\documentclass{article}

\usepackage{arxiv}

\usepackage[utf8]{inputenc}
\usepackage[T1]{fontenc}   
\usepackage{hyperref}      
\usepackage{url}           
\usepackage{booktabs}      
\usepackage{amsfonts}      
\usepackage{nicefrac}      
\usepackage{microtype}     
\usepackage{lipsum}		
\usepackage{graphicx}
\usepackage[square,numbers]{natbib}
\usepackage{doi}

\usepackage{amsmath}
\usepackage{amssymb}
\usepackage{array}
\usepackage{caption}

\title{Intelligent Scientific Literature Explorer Using Machine Learning (ISLE)}

\author{
 \href{https://orcid.org/0009-0000-8882-145X}{\includegraphics[scale=0.06]{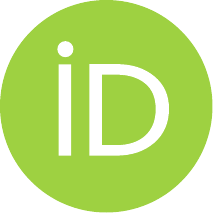}\hspace{1mm}Sina Jani}\thanks{Corresponding author.} \\
Computer Engineering Department \\
  Allameh Tabatabaei University \\
  Tehran, Iran\\
    \texttt{jani\_s@atu.ac.ir} \\
\And
Arman Heidari\\
Computer Engineering Department \\
 Allameh Tabatabaei University \\
Tehran, Iran\\
\texttt{arman\_heidari@atu.ac.ir} \\
\AND
Amirmohammad Anvari\\
Computer Engineering Department \\
Allameh Tabatabaei University \\
 Tehran, Iran\\
\texttt{am\_anvari@atu.ac.ir} \\
\And
Zahra Rahimi\\
Computer Engineering Department \\
 Allameh Tabatabaei University \\
 Tehran, Iran\\
\texttt{za.rah@atu.ac.ir} \\
}

\renewcommand{\headeright}{}
\renewcommand{\undertitle}{}

\begin{document}
\maketitle

\begin{abstract}
 The rapid acceleration of scientific publishing has created substantial challenges for researchers attempting to discover, contextualize, and interpret relevant literature. Traditional keyword-based search systems provide limited semantic understanding, while existing AI-driven tools typically focus on isolated tasks such as retrieval, clustering, or bibliometric visualization. This paper presents an integrated system for scientific literature exploration that combines large-scale data acquisition, hybrid retrieval, semantic topic modeling, and heterogeneous knowledge graph construction. The system builds a comprehensive corpus by merging full-text data from arXiv with structured metadata from OpenAlex. A hybrid retrieval architecture fuses BM25 lexical search with embedding-based semantic search using Reciprocal Rank Fusion. Topic modeling is performed on retrieved results using BERTopic or non-negative matrix factorization depending on computational resources. A knowledge graph unifies papers, authors, institutions, countries, and extracted topics into an interpretable structure. The system provides a multi-layered exploration environment that reveals not only relevant publications but also the conceptual and relational landscape surrounding a query. Evaluation across multiple queries demonstrates improvements in retrieval relevance, topic coherence, and interpretability. The proposed framework contributes an extensible foundation for AI-assisted scientific discovery.
\end{abstract}


\section{Introduction}
The accelerating expansion of global scientific output has made it increasingly difficult for researchers to identify, interpret, and contextualize relevant literature. Bibliometric analyses demonstrate that the number of cited references in scholarly communication has grown exponentially at an annual rate of approximately 8--9\% since the mid-twentieth century \cite{bornmann2015growth}, resulting in an ever-denser and more fragmented research landscape. This growth has amplified the limitations of traditional academic search systems, many of which rely primarily on keyword-based retrieval and provide limited transparency, metadata control, or reproducibility. Comparative evaluations of academic search engines show that widely used platforms often fail to satisfy essential requirements for systematic discovery, advanced filtering, or semantic interpretation \cite{gusenbauer2020academic}, and even high-coverage systems such as Google Scholar exhibit inconsistent indexing and limited query expressiveness \cite{gusenbauer2019google}.

Recent benchmarks highlight that scientific literature search differs substantially from generic information retrieval: research queries frequently require capturing conceptual, methodological, and citation-based relationships. Neural embedding--based retrieval significantly surpasses BM25 for scientifically meaningful queries, while lexical search remains valuable for precise terminology matching, motivating hybrid retrieval architectures \cite{ajith2024litsearch}. At the same time, advances in large-scale scholarly metadata integration have produced comprehensive scientific knowledge graphs such as OpenAlex \cite{priem2022openalex} and PubGraph \cite{ahrabian2023pubgraph}, which link papers, authors, institutions, venues, and concepts at global scale. However, these resources are primarily designed for static metadata aggregation rather than dynamic, query-conditioned exploration. Complementary work on academic and research knowledge graphs demonstrates the value of structured representations for scholarly understanding \cite{du2022academic, zloch2025research, peng2023knowledge}, yet these systems typically do not integrate retrieval, topic modeling, and interactive graph generation into a unified pipeline.

Parallel progress in representation learning has transformed scientific document understanding. Citation-informed models such as SPECTER \cite{cohan2020specter} and sentence-embedding architectures based on Siamese BERT networks \cite{reimers2019sentence} provide semantically meaningful vector representations of scientific texts. Lightweight transformer variants, such as MiniLM \cite{song2020minilm}, enable efficient large-scale indexing and clustering, making them suitable for corpus-level retrieval systems. For thematic analysis, neural topic modeling approaches such as BERTopic \cite{grootendorst2022bertopic} consistently outperform classical models like LDA \cite{blei2003lda} and NMF \cite{lee1999nmf} in coherence when applied to scientific abstracts, particularly when combined with transformer-derived embeddings \cite{chagnon2024benchmarking, sharifianattar2023topic}. Yet topic modeling pipelines are generally decoupled from retrieval processes and do not integrate with graph-based representation frameworks.

Hybrid retrieval research has demonstrated that combining lexical and semantic signals---often using Reciprocal Rank Fusion---yields superior performance and robustness across heterogeneous query types \cite{bruch2023fusion, elastic2023hybrid, opensearch2024hybrid}. These findings underscore the need for integrated systems that leverage both symbolic and embedding-based information to support deep exploration of scientific literature.

To address these challenges, this paper introduces the Intelligent Scientific Literature Explorer (ISLE), an end-to-end machine-learning--driven system for scientific literature exploration. ISLE constructs a large, unified corpus by merging full-text data from arXiv with structured metadata from OpenAlex, enabling comprehensive entity coverage spanning papers, authors, institutions, countries, citations, and citing works. Retrieval is conducted through a hybrid retrieval architecture that fuses BM25 lexical search with semantic search using all-MiniLM-L6-v2 embeddings via Reciprocal Rank Fusion, integrating the complementary strengths of both paradigms \cite{song2020minilm, bruch2023fusion, elastic2023hybrid, opensearch2024hybrid}. ISLE then applies topic modeling on the retrieved subset using either BERTopic \cite{grootendorst2022bertopic} or NMF \cite{lee1999nmf}, depending on computational resources, to generate interpretable thematic clusters. Finally, the system constructs a heterogeneous knowledge graph linking retrieved papers to their associated authors, institutions, countries, and extracted topics, providing users with a structured, multi-layered view of the conceptual and relational landscape surrounding a query.

Through the integration of large-scale data acquisition, hybrid retrieval, transformer-based semantic modeling, topic extraction, and knowledge graph construction, ISLE moves beyond traditional document-centric search tools and offers a comprehensive environment for AI-assisted scientific discovery. The system provides enhanced retrieval relevance, coherent topic structure, and interpretable relational insights, addressing key limitations in current scholarly exploration technologies.

The remainder of this paper is structured as follows. Section \ref{sec2} reviews related work in academic search systems, hybrid retrieval, topic modeling, and scientific knowledge graphs. Section \ref{sec3} details the proposed ISLE framework, including data acquisition and integration, hybrid lexical–semantic retrieval, resource-aware topic modeling, dynamic knowledge graph construction, and the end-to-end system architecture. Section \ref{sec4} presents an evaluation of the system, encompassing corpus characteristics, retrieval behavior, and a qualitative case study. Finally, Section \ref{sec5} discusses the implications and limitations of the proposed approach and concludes with directions for future work.

\section{Related Works}\label{sec2}
Research on scientific literature exploration spans several interconnected areas, including academic search systems, hybrid retrieval, semantic embedding models, topic modeling, and scientific knowledge graphs. This section reviews the most relevant developments and positions the contributions of the Intelligent Scientific Literature Explorer (ISLE) within this landscape.

\subsection{Academic Search Systems}

Academic search engines such as Google Scholar, Web of Science, and Scopus remain the primary entry points for literature discovery, yet they exhibit substantial limitations. Systematic evaluations reveal that many scholarly search systems lack the transparency, reproducibility, and metadata control required for rigorous research workflows \cite{gusenbauer2020academic}. Coverage analyses further demonstrate that although Google Scholar indexes a much broader range of documents than subscription-based platforms, its indexing methodology is opaque and inconsistent, and result sets are unstable across repeated queries \cite{gusenbauer2019google}. These systems also provide minimal semantic understanding, relying predominantly on lexical matching, which constrains their ability to handle queries involving conceptual relationships, methodological dependencies, or latent topic structures. The emergence of domain-specific retrieval benchmarks confirms that conventional search engines are inadequate for complex scientific queries requiring deeper semantic reasoning \cite{ajith2024litsearch}.

\subsection{Information Retrieval}

Advances in representation learning have led to substantial improvements in semantic search for scientific documents. Citation-aware architectures such as SPECTER \cite{cohan2020specter} generate document representations that capture citation semantics, improving performance in tasks such as co-citation prediction and paper recommendation. Likewise, Sentence-BERT \cite{reimers2019sentence} and its successors enable efficient similarity computation by producing fixed-length embeddings through Siamese or triplet architectures. Lightweight transformer distillation techniques such as MiniLM \cite{song2020minilm} preserve semantic quality while enabling large-scale retrieval and clustering.

Although dense retrieval significantly outperforms BM25 on conceptually rich scientific queries \cite{ajith2024litsearch, mori2025performance}, lexical search remains competitive for exact term matching and domain-specific keywords \cite{formal2021match}. This complementarity has motivated hybrid retrieval approaches that fuse lexical and embedding-based similarity signals. Rank-based fusion---particularly Reciprocal Rank Fusion (RRF)---consistently yields robust gains across heterogeneous scoring distributions \cite{bruch2023fusion}. Hybrid retrieval implementations in modern IR frameworks such as Elasticsearch and OpenSearch demonstrate similar performance improvements, validating the practical value of combined lexical--semantic pipelines \cite{elastic2023hybrid, opensearch2024hybrid}. These developments inform ISLE's hybrid retrieval architecture, which integrates BM25 with all-MiniLM-L6 embeddings using RRF.

\subsection{Topic Modeling}

Topic modeling provides an important mechanism for summarizing and structuring large scientific corpora. Classical models such as Latent Dirichlet Allocation (LDA) \cite{blei2003lda} and Non-Negative Matrix Factorization (NMF) \cite{lee1999nmf} remain widely used due to computational efficiency and interpretability. However, empirical evaluations on scientific text show that classical models often exhibit lower topic coherence compared to neural topic models, especially when dealing with high-dimensional embeddings. BERTopic \cite{grootendorst2022bertopic}, which combines transformer-derived embeddings with density-based clustering and a class-based TF-IDF representation, has emerged as a strong baseline for scientific abstracts. Comparative studies across scientific disciplines find that BERTopic produces more coherent and semantically meaningful topics than LDA or NMF \cite{chagnon2024benchmarking}, particularly when using BERT-based or MiniLM embeddings \cite{sharifianattar2023topic}. Nevertheless, large-scale neural topic modeling requires GPU resources and is typically decoupled from retrieval systems, limiting its application in interactive literature exploration. ISLE addresses this gap by performing query-specific topic modeling on retrieved documents using BERTopic when GPU resources are available and NMF as a CPU-efficient alternative.

\subsection{Scientific Knowledge Graphs}

Knowledge graphs have become central to representing scholarly entities and their interdependencies. OpenAlex \cite{priem2022openalex} provides a large-scale, openly accessible scientific knowledge graph that integrates metadata from Crossref, PubMed, and other sources, covering works, authors, institutions, venues, and hierarchical research concepts. PubGraph \cite{ahrabian2023pubgraph} extends this paradigm by constructing a temporal scientific knowledge graph integrating data from Wikidata, OpenAlex, and Semantic Scholar, supporting complex downstream tasks such as link prediction and temporal analysis. Complementary efforts such as the Academic Paper Knowledge Graph (APKG) \cite{du2022academic} aim to model the internal semantic structure of individual documents, while recent surveys on research knowledge graphs categorize existing systems and highlight their role in improving the FAIRness of scholarly metadata \cite{zloch2025research, peng2023knowledge}. Despite this progress, existing scientific knowledge graphs are static and global in scope---they are not designed to dynamically incorporate retrieval results, extracted topics, or user-specific contextual information.

\subsection{Positioning of ISLE}

While prior work has advanced retrieval models, topic modeling techniques, and large-scale scholarly knowledge graphs independently, few systems integrate these components into a unified workflow. ISLE differentiates itself by combining:

\begin{enumerate}
    \item large-scale data acquisition from arXiv and OpenAlex;
    \item a hybrid lexical--semantic retrieval pipeline leveraging BM25 and all-MiniLM-L6-v2 embeddings fused via RRF;
    \item dynamic topic modeling tailored to retrieved documents using BERTopic or NMF ; and
    \item construction of a heterogeneous knowledge graph linking papers, topics, authors, institutions, countries, and citation relationships.
\end{enumerate}

This integration enables a multi-layered, interpretable exploration of the scientific landscape, addressing the methodological and functional limitations identified in prior work.

\section{Methodology}\label{sec3}

This section presents the methodological framework underlying ISLE, detailing the end-to-end design and implementation of a query-conditioned scientific literature exploration system. The methodology integrates large-scale data acquisition, hybrid retrieval, resource-aware topic modeling, and dynamic knowledge graph construction into a unified analytical pipeline. Each component is designed to balance semantic expressiveness, computational efficiency, and scalability, while remaining responsive to user-specific information needs. Rather than operating on a static global corpus, ISLE constructs query-conditioned representations that enable focused retrieval, thematic abstraction, and relational reasoning over scientific knowledge. Symbols are locally scoped to each subsection unless otherwise stated.

\subsection{Data Acquisition and Integration}

To build a comprehensive and interlinked scholarly corpus, ISLE integrates data from two complementary sources: arXiv and OpenAlex. A full snapshot from both sources---covering all available records from their inception through August 2025---was collected. This temporal range ensures that the corpus reflects the historical evolution and the current state of scientific research.

The choice of sources is motivated by their orthogonal strengths. arXiv provides timely access to full-text preprints, abstracts, and subject categories, making it particularly suitable for content-rich semantic analysis. OpenAlex, by contrast, offers structured and clean metadata, including citation networks, author affiliations, institutional hierarchies, and country-level attributes \cite{priem2022openalex}.

To maintain computational feasibility while preserving thematic relevance, the acquisition pipeline filters the dataset to include only records belonging to the Computer Science (cs.*) and Physics (physics.*) categories of arXiv. These two domains collectively represent a large portion of high-velocity research activity and include substantial overlap in machine learning--related literature, which is essential for ISLE's semantic retrieval and topic modeling components.

\subsubsection{Data Schema Design}

The integrated records were stored in a relational database (PostgreSQL) to enforce referential integrity, normalize heterogeneous metadata, and support efficient join queries across entities such as papers, authors, institutions, and citations \cite{han2011data}.

The final schema centers around three core tables---papers, authors, and citations---with additional linking tables used to support many-to-many relationships.

\begin{table}[h]
\centering
\begin{tabular}{>{\ttfamily}l p{10cm}}
\toprule
\normalfont\textbf{Field} & \textbf{Description} \\
\midrule
arxiv\_id & Unique identifier assigned by arXiv. \\
title & Title of the publication. \\
publication\_date & Date of public release (from OpenAlex when available). \\
submitted\_date & Original submission timestamp from arXiv. \\
doi & Digital Object Identifier for cross-platform referencing. \\
subject & arXiv subject classification (e.g., cs.CV, physics.optics). \\
\bottomrule
\end{tabular}
\vspace{0.5em}
\caption{Papers Table}
\label{tab:papers}
\end{table}

This table serves as the backbone of the relational model, storing canonical metadata for each publication.

\begin{table}[h]
\centering
\begin{tabular}{>{\ttfamily}l p{10cm}}
\toprule
\normalfont\textbf{Field} & \textbf{Description} \\
\midrule
author\_id & Internal unique identifier. \\
related\_institutions & List of institution identifiers mapped from OpenAlex. \\
related\_countries & Countries corresponding to affiliated institutions. \\
\bottomrule
\end{tabular}
\vspace{0.5em}
\caption{Authors Table}
\label{tab:authors}
\end{table}

Affiliation data is resolved through OpenAlex's institution entities, enabling geographic and organizational aggregation during knowledge graph construction.

\begin{table}[h]
\centering
\begin{tabular}{>{\ttfamily}l p{10cm}}
\toprule
\normalfont\textbf{Field} & \textbf{Description} \\
\midrule
citing\_paper\_id & Paper initiating the citation. \\
cited\_paper\_id & Paper being referenced. \\
\bottomrule
\end{tabular}
\vspace{0.5em}
\caption{Citations Table}
\label{tab:citations}
\end{table}

This table encodes citation links and is optimized for graph traversal queries and downstream analytics such as bibliographic coupling, co-citation analysis, and lineage discovery.

\subsubsection{Search Index Preparation}
To enable high-performance hybrid retrieval, the corpus was indexed in a dedicated search engine (Elasticsearch). This secondary index bridges the user's query space with the document corpus, supporting both lexical matching and semantic similarity search.

Each document is vectorized using the all-MiniLM-L6-v2 model \cite{song2020minilm}, producing a 384-dimensional embedding from the concatenation of a paper's title and abstract. To support both BM25 lexical retrieval and dense vector search, a custom index mapping was implemented. Key textual fields---such as title and abstract---utilize the built-in English analyzer, which applies stemming and stop-word removal. Author names and institution fields include both tokenized and keyword-level representations to support aggregation and filtering.

\begin{table}[h]
\centering
\small
\begin{tabular}{>{\ttfamily}l l l p{5cm}}
\toprule
\normalfont\textbf{Field} & \textbf{Type} & \textbf{Analyzer} & \textbf{Description} \\
\midrule
paper\_id & keyword & --- & Unique paper identifier (linked to PostgreSQL). \\
title & text & english & Paper title, includes raw keyword subfield. \\
abstract & text & english & Full abstract text. \\
authors & text & standard & Tokenized author names, with keyword subfield. \\
countries & keyword & --- & Countries of author affiliations. \\
institutions & text & standard & Institutional names, with keyword subfield. \\
embedding & dense\_vector & --- & 384-dim MiniLM embedding of title + abstract. \\
\bottomrule
\end{tabular}
\vspace{0.5em}
\caption{Elasticsearch Index Mapping for the Scientific Corpus}
\label{tab:elasticsearch}
\end{table}

This hybrid indexing strategy provides the foundation for ISLE's Reciprocal Rank Fusion (RRF) retrieval framework, combining lexical and semantic signals in downstream queries.

\subsection{Hybrid Retrieval Architecture}

The retrieval subsystem of ISLE is designed to meet the diverse information needs of scientific search, where a single query may require both precise lexical matching and deep conceptual understanding. To address this dual requirement, ISLE implements a hybrid retrieval architecture that integrates lexical and semantic search pipelines. Upon receiving a user query, the system generates a dense vector representation of the query, executes a BM25-based lexical search and a dense vector--based semantic search in parallel, and then fuses their ranked outputs using Reciprocal Rank Fusion (RRF) to produce a unified ranking.

This design is motivated by the complementary strengths of the two retrieval paradigms. Dense neural retrieval consistently outperforms BM25 on conceptually rich scientific queries by capturing latent relationships beyond surface-level text \cite{ajith2024litsearch}, whereas lexical search remains superior for matching precise technical terms, acronyms, formulas, and named entities frequently encountered in scientific writing \cite{bruch2023fusion, mitra2018nir}. Systems that rely exclusively on either lexical or semantic approaches exhibit significant blind spots, particularly on heterogeneous query types. Hybrid retrieval frameworks that combine both signals---especially through rank-based fusion---repeatedly demonstrate improved robustness and relevance across diverse tasks \cite{elastic2023hybrid, opensearch2024hybrid}. By leveraging RRF to integrate these complementary signals, ISLE delivers retrieval results that simultaneously reflect precise terminology and high-level conceptual similarity.

\subsubsection{Lexical Retrieval}

The lexical retrieval component captures exact or near-exact textual matches, which remain essential for scientific literature where specialized terminology and mathematically grounded expressions carry precise meaning \cite{manning2008introduction}. ISLE implements the BM25 algorithm, which extends classical TF--IDF by accounting for term-frequency saturation and document-length normalization, enabling reliable relevance estimation in scientific corpora.

The BM25 relevance score of a document $D$ given a query $Q$ containing terms $q_1,\ldots,q_n$ is calculated as shown in Equation~\ref{eq:bm25}:

\begin{equation}
\text{Score}(D,Q) = \sum_{i=1}^{n} \text{IDF}(q_i) \cdot \frac{f(q_i, D) \cdot (k_1+1)}{f(q_i, D) + k_1 \cdot \left(1-b + b \cdot \frac{|D|}{\text{avgdl}}\right)}
\label{eq:bm25}
\end{equation}

where:
\begin{itemize}
    \item $f(q_i, D)$ = frequency of term $q_i$ in document $D$
    \item $|D|$ = length of document $D$ in words
    \item $\text{avgdl}$ = average document length in the corpus
    \item $k_1$ and $b$ = free parameters (typically $k_1 \in [1.2, 2.0]$ and $b=0.75$)
    \item $\text{IDF}(q_i)$ = inverse document frequency of term $q_i$
\end{itemize}

To optimize retrieval for scientific discourse, ISLE employs a field-aware ranking strategy that boosts exact and phrase-level matches, while incorporating fuzzy matching to handle morphological variations and minor misspellings. Scientific information needs often require both precision (capturing specific technical terms) and recall (accounting for terminological variation) \cite{ajith2024litsearch, manning2008introduction}.

The top-$k$ documents from the lexical subsystem are forwarded to the fusion stage, ensuring that terminological precision is preserved within the overall hybrid retrieval framework.

\subsubsection{Semantic Retrieval}

The semantic retrieval path addresses the principal limitation of lexical search---its inability to retrieve documents that use different wording to convey conceptually similar ideas. ISLE implements dense vector retrieval, mapping both queries and documents into a shared high-dimensional embedding space where semantic proximity reflects conceptual relatedness.

User queries are encoded using the all-MiniLM-L6-v2 sentence transformer model, the same model used during corpus preprocessing. Embedding both queries and documents using an identical encoder maintains geometric consistency within the embedding space. This transformer variant balances semantic fidelity, speed, and computational efficiency, making it well-suited for interactive retrieval environments \cite{song2020minilm}.

Semantic similarity is computed using cosine similarity, a standard metric for evaluating directional alignment between high-dimensional vectors (Equation~\ref{eq:cosine}):

\begin{equation}
\text{Similarity}(Q,D) = \cos\theta = \frac{Q \cdot D}{\|Q\| \|D\|}
\label{eq:cosine}
\end{equation}

where $\theta$ is the angle between the query vector $Q$ and document vector $D$.

A $k$-nearest neighbors (KNN) search retrieves the most semantically similar documents. This enables the system to identify relevant works even when they lack direct lexical overlap with the query---for example, when two papers describe the same concept using different terminology, refer to related methodologies, or are part of the same scientific subfield.

The top-$k$ semantically relevant documents are passed to the fusion stage to ensure that conceptual relevance is well represented in the final ranking.

\subsubsection{Reciprocal Rank Fusion}

To unify the lexical and semantic retrieval signals, ISLE employs Reciprocal Rank Fusion (RRF), a robust and score-independent rank aggregation method particularly suitable for hybrid search systems. Unlike score-based fusion methods, RRF does not require normalization across disparate scoring scales, making it ideal for combining BM25 and cosine-similarity rankings \cite{bruch2023fusion, opensearch2024hybrid}.

For each document $D$ appearing in either the lexical ranking $R_L$ or semantic ranking $R_S$, the fused score is computed as (Equation~\ref{eq:rrf}):

\begin{equation}
\text{RRF}(D) = \sum_{r \in \{R_L, R_S\}} \frac{1}{k + \text{rank}_r(D)}
\label{eq:rrf}
\end{equation}

where:
\begin{itemize}
    \item $\text{rank}_r(D)$ = rank position of document $D$ in ranking $r$ ($\text{rank}_r(D) = \infty$ if $D$ does not appear in $r$)
    \item $k$ = a smoothing constant typically set to 60, which ensures reasonable scores for documents that appear deep in the rankings \cite{bruch2023fusion, cormack2009rrf}
\end{itemize}

This approach effectively promotes documents that achieve high rankings in both retrieval systems while still preserving relevant documents that are highly ranked by only one system. The method is particularly valuable in scientific literature search where different query aspects may be better captured by lexical or semantic approaches \cite{opensearch2024hybrid}.

The resulting fused list is sorted by descending RRF score, and the top-$N$ documents---where $N$ is a configurable system parameter---are forwarded to ISLE's downstream modules for topic modeling and knowledge graph construction.

\subsection{Topic Modeling}

To support exploratory search, trend analysis, and thematic structuring of the scholarly corpus, ISLE integrates a dual-path topic modeling framework that balances computational efficiency with semantic fidelity. The system employs two complementary approaches: a classical statistical pipeline based on Non-Negative Matrix Factorization (NMF) and a neural topic modeling pipeline based on BERTopic. ISLE can operate effectively across heterogeneous deployment environments, from resource-constrained systems to high-performance infrastructures.

Across both configurations, topic induction operates over concatenated title and abstract fields, which provide high signal-to-noise ratios for scientific domain characterization without requiring full-text ingestion \cite{tekin2022abstract}. Prior to modeling, documents are normalized and filtered to remove excessively short or malformed abstracts.

\subsubsection{NMF-Based Topic Modeling}

For resource-constrained environments, ISLE employs a classical semantic decomposition technique rooted in Non-Negative Matrix Factorization (NMF) \cite{lee1999nmf}. Each document is represented using TF--IDF vectors, where the TF--IDF weight for term $t$ in document $d$ is defined as (Equation~\ref{eq:tfidf}):

\begin{equation}
\text{TF-IDF}(t,d) = \text{tf}(t,d) \times \log\left(\frac{N}{\text{df}(t)}\right)
\label{eq:tfidf}
\end{equation}

where:
\begin{itemize}
    \item $\text{tf}(t,d)$ = frequency of term $t$ in document $d$
    \item $\text{df}(t)$ = number of documents containing $t$
    \item $N$ = total number of documents
\end{itemize}

Given the document-term matrix $X \in \mathbb{R}_+^{m \times n}$, NMF decomposes $X$ into two low-rank non-negative matrices (Equation~\ref{eq:nmf}):

\begin{equation}
X \approx WH
\label{eq:nmf}
\end{equation}

where:
\begin{itemize}
    \item $W \in \mathbb{R}_+^{m \times k}$ contains document-topic weights
    \item $H \in \mathbb{R}_+^{k \times n}$ contains topic-term distributions
    \item $k$ is the number of topics
\end{itemize}

This optimization is typically expressed as (Equation~\ref{eq:nmf_opt}):

\begin{equation}
\arg\min_{W, H \geq 0} \|X - WH\|_F^2
\label{eq:nmf_opt}
\end{equation}

To select $k$, ISLE performs a coherence-driven model selection sweep. Topic coherence is computed using normalized PMI (NPMI) over the top-$n$ topic words (Equation~\ref{eq:npmi}):

\begin{equation}
\text{NPMI}(w_i,w_j) = \frac{\log \frac{P(w_i,w_j)}{P(w_i)P(w_j)}}{\log P(w_i,w_j)}
\label{eq:npmi}
\end{equation}

The optimal $k$ maximizes mean topic coherence \cite{newman2010evaluation, roder2015coherence}.

\subsubsection{BERTopic-Based Neural Topic Modeling}

When computational resources allow, ISLE integrates BERTopic, a transformer-based framework combining contextual embeddings, non-linear manifold learning, density-based clustering, and class-based TF--IDF refinement \cite{grootendorst2022bertopic}. Each document $d$ is encoded into a vector (Equation~\ref{eq:bertopic_encode}):

\begin{equation}
\mathbf{e}_d = f_\theta(d)
\label{eq:bertopic_encode}
\end{equation}

where $f_\theta$ denotes the Qwen3-Embedding-0.6B encoding model \cite{zhang2025qwen3}.

Dimensionality reduction is performed via UMAP, which learns a low-dimensional manifold preserving local neighborhood structure \cite{maaten2008tsne, hinton2006dimensionality}. The transformation is expressed as (Equation~\ref{eq:umap}):

\begin{equation}
\mathbf{z}_d = \text{UMAP}(\mathbf{e}_d) \quad \text{where} \quad \mathbf{z}_d \in \mathbb{R}^p \text{ and } p \ll 1024
\label{eq:umap}
\end{equation}

Clusters are then identified using HDBSCAN, which extracts variable-density clusters by maximizing stability (Equation~\ref{eq:hdbscan}):

\begin{equation}
C = \arg\max_C \text{Stability}(C)
\label{eq:hdbscan}
\end{equation}

Topic words for each cluster are computed via class-based TF--IDF (c-TF-IDF) as shown in Equation~\ref{eq:ctfidf}:

\begin{equation}
\text{CTF-IDF}(t,c) = \frac{\text{tf}(t,c)}{|c|} \times \log\left(\frac{N}{\text{df}(t)}\right)
\label{eq:ctfidf}
\end{equation}

where $|c|$ = total token count of the cluster concatenation.

This neural approach yields semantically coherent and automatically labeled high-level research themes, capturing conceptual similarities beyond lexical overlap \cite{grootendorst2022bertopic, egger2022topic, mutsaddi2025bertopic, mersha2025semantic}.

\subsubsection{Topic Evaluation and Integration}

Both pipelines are evaluated using standard coherence metrics such as (Equation~\ref{eq:coherence}):

\begin{equation}
\text{Coherence}_{c_v} = \frac{1}{\binom{n}{2}} \sum_{i < j} \text{NPMI}(w_i,w_j)
\label{eq:coherence}
\end{equation}

ensuring interpretability and thematic consistency \cite{newman2010evaluation, roder2015coherence}.

Topic distributions serve downstream ISLE capabilities including interactive exploration, temporal evolution analysis, and integration into the knowledge graph navigation system.

\subsection{Knowledge Graph Construction}

To enable structured exploration, relational reasoning, and contextual navigation over retrieved scientific literature, ISLE constructs a dynamic scholarly knowledge graph tailored to each user query. Unlike static, corpus-wide research knowledge graphs \cite{ahrabian2023pubgraph, du2022academic, zloch2025research}, the ISLE knowledge graph is instantiated on demand, conditioned on the set of documents returned by the hybrid retrieval and topic modeling pipelines. The resulting knowledge graph reflects the user's immediate information need while remaining computationally tractable and semantically focused.

Given a user query $q$, the retrieval subsystem returns a ranked document set:

\begin{equation}
\mathcal{D}_q = \{d_1, d_2, \ldots, d_n\}
\label{eq:doc_set}
\end{equation}

using the hybrid retrieval framework described earlier. Topic modeling is then performed over $\mathcal{D}_q$, producing topic assignments and probabilities that directly inform graph construction. The final knowledge graph is built exclusively from this query-conditioned subset, rather than the full corpus.

\subsubsection{Graph Definition}

The ISLE knowledge graph is formally defined as a typed, directed, attributed multigraph (Equation~\ref{eq:graph}):

\begin{equation}
\mathcal{G}_q = (V_q, E_q)
\label{eq:graph}
\end{equation}

where $V_q$ denotes the set of nodes and $E_q$ the set of directed edges induced by query $q$. Multiple edges between node pairs are permitted to represent distinct semantic relations.

\subsubsection{Node Types and Attributes}

The node set $V_q$ consists of heterogeneous entities (Equation~\ref{eq:nodes}):

\begin{equation}
V_q = V_P \cup V_A \cup V_I \cup V_C \cup V_T \cup V_Y
\label{eq:nodes}
\end{equation}

where:
\begin{itemize}
    \item $V_P$: paper nodes
    \item $V_A$: author nodes
    \item $V_I$: institution nodes
    \item $V_C$: country nodes
    \item $V_T$: topic nodes
    \item $V_Y$: temporal (year) nodes
\end{itemize}

Each paper node $p \in V_P$ is associated with attributes:

\begin{equation}
p = \{\text{title}, \text{publication year}, \text{citation count}, \text{topic}, \pi_{p,t}\}
\label{eq:paper_node}
\end{equation}

where $\pi_{p,t}$ denotes the probability of paper $p$ belonging to topic $t$, as produced by the topic modeling stage.

Topic nodes $t \in V_T$ encapsulate:

\begin{equation}
t = \{\text{topic id}, \text{keywords}, \text{document count}\}
\label{eq:topic_node}
\end{equation}

where keywords are extracted using class-based TF--IDF in the neural setting \cite{grootendorst2022bertopic} or term-weight analysis in the NMF setting \cite{lee1999nmf}.

\subsubsection{Edge Types and Semantics}

Edges encode explicit scholarly relationships and are represented as labeled, directed relations (Equation~\ref{eq:edges}):

\begin{equation}
E_q = E_{\text{auth}} \cup E_{\text{aff}} \cup E_{\text{geo}} \cup E_{\text{temp}} \cup E_{\text{topic}} \cup E_{\text{cite}}
\label{eq:edges}
\end{equation}

where:
\begin{itemize}
    \item Authorship: $(a,p) \in E_{\text{auth}}$ if author $a$ authored paper $p$
    \item Institutional affiliation: $(p,i) \in E_{\text{aff}}$ if paper $p$ is affiliated with institution $i$
    \item Geographic attribution: $(p,c) \in E_{\text{geo}}$ if paper $p$ originates from country $c$
    \item Temporal publication: $(p,y) \in E_{\text{temp}}$ if paper $p$ was published in year $y$
    \item Topic assignment: $(p,t) \in E_{\text{topic}}$ with weight $\pi_{p,t}$
    \item Citation relationships: $(p_i,p_j) \in E_{\text{cite}}$ if paper $p_i$ cites paper $p_j$
\end{itemize}

Citation edges follow the standard bibliographic direction, forming a directed citation graph suitable for lineage analysis, co-citation studies, and traversal-based analytics \cite{ahrabian2023pubgraph, du2022academic, peng2023knowledge}.

\subsubsection{Graph-Topological Impact Aggregation}

Entity-level impact is computed using pure graph topology, rather than metadata propagation. For a paper $p$, citation impact corresponds to its indegree in the citation subgraph (Equation~\ref{eq:paper_impact}):

\begin{equation}
\text{Impact}(p) = \deg_{E_{\text{cite}}}^{-}(p)
\label{eq:paper_impact}
\end{equation}

For higher-order entities such as authors, institutions, and countries, impact is derived as the cumulative citation indegree of their associated papers (Equation~\ref{eq:entity_impact}):

\begin{equation}
\text{Impact}(x) = \sum_{p \in \mathcal{P}(x)} \deg_{E_{\text{cite}}}^{-}(p)
\label{eq:entity_impact}
\end{equation}

where $\mathcal{P}(x) \subseteq V_P$ denotes the set of papers connected to entity $x$ via authorship, affiliation, or geographic edges. This formulation preserves strict bibliographic semantics and enables consistent comparison across entity types \cite{bornmann2015growth, ahrabian2023pubgraph}.

\subsubsection{Dynamic Scale and Query Dependence}

Because graph construction is conditioned on $\mathcal{D}_q$, the size and density of $\mathcal{G}_q$ vary with the user query. Empirically, ISLE produces graphs ranging from approximately 40,000 paper nodes to 800,000 total edges per query, depending on retrieval breadth and topic granularity. This query-scoped design significantly reduces noise, improves interpretability, and enables interactive exploration without the computational overhead of maintaining a monolithic global knowledge graph \cite{zloch2025research, peng2023knowledge}.

The resulting knowledge graph integrates hybrid retrieval results, topic modeling outputs, citation topology, and relational metadata. It supports downstream applications such as topic-centric navigation, influence analysis, temporal trend discovery, and structured recommendation, forming a core analytical layer within the ISLE platform.

\subsection{System Architecture and End-to-End Pipeline}

ISLE is implemented as a modular, query-conditioned architecture that orchestrates retrieval, topic modeling, and relational analysis into a single end-to-end execution pipeline. The system constructs query-conditioned analytical artifacts---rather than relying on static, corpus-wide structures---thereby enabling focused exploration, improved interpretability, and scalable execution. Figure~\ref{fig:query_processing} and Figure~\ref{fig:topic_kg} illustrate the two principal stages of this pipeline: (i) query processing and hybrid retrieval, and (ii) resource-aware topic modeling followed by dynamic knowledge graph construction.

\subsubsection{Query Preprocessing and Hybrid Retrieval}

The pipeline begins with user query ingestion, where the input text is normalized to ensure consistency across downstream components. Preprocessing includes standard text normalization steps such as lowercasing, whitespace normalization, and punctuation handling, producing a canonical representation of the query suitable for both lexical and semantic processing.

\begin{figure}[h]
\centering
\includegraphics[width=0.8\textwidth]{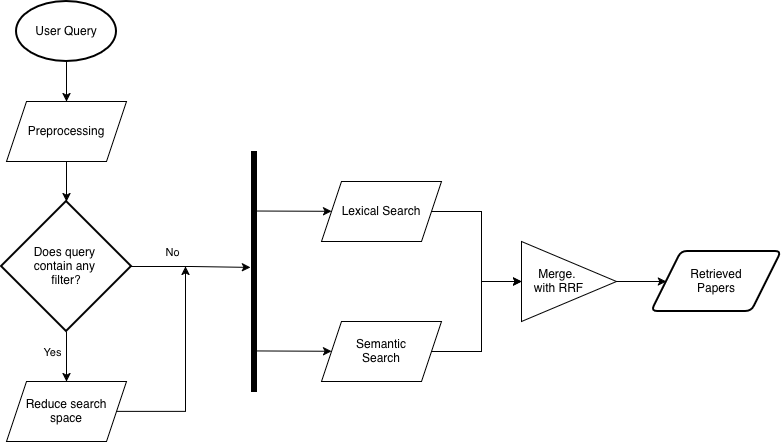}
\caption{Query processing stage}
\label{fig:query_processing}
\end{figure}

In addition to free-text input, ISLE supports structured filters specified by the user, including publication year ranges, author names, institutional affiliations, and country attributes. When present, these constraints are applied prior to retrieval to reduce the effective search space. This early filtering stage improves computational efficiency and precision by limiting candidate documents to those satisfying explicit user criteria, while preserving the full expressive power of the hybrid retrieval framework.

Following preprocessing and optional filtering, the normalized query is dispatched to the hybrid retrieval subsystem. Lexical and semantic searches are executed in parallel over the filtered corpus subset, and their ranked outputs are consolidated using Reciprocal Rank Fusion (RRF), as detailed in Section~3.2.3 and formalized in Equation~\ref{eq:rrf}. This stage yields a ranked set of retrieved papers tailored to both the semantic intent of the query and any user-imposed constraints. Figure~\ref{fig:query_processing} depicts the overall control flow, highlighting the parallel execution and subsequent fusion of retrieval signals.

\subsubsection{Topic Modeling and Dynamic Knowledge Graph}

\begin{figure}[h]
    \centering
    \includegraphics[width=0.8\textwidth]{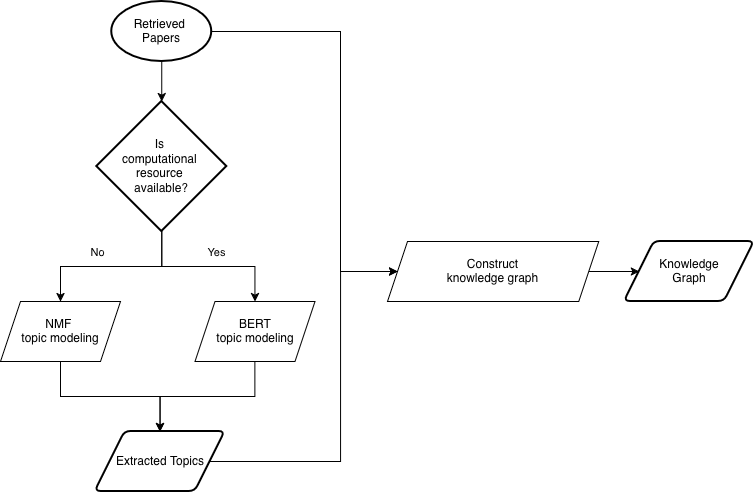}
    \caption{Resource-aware topic modeling followed by dynamic knowledge graph construction}
    \label{fig:topic_kg}
\end{figure}

The outputs of retrieval and topic modeling jointly inform the construction of a query-conditioned knowledge graph, as described in Section~3.4. The knowledge graph is instantiated dynamically for each query, incorporating only entities and relationships present in the retrieved document set and their associated metadata. As illustrated in Figure~\ref{fig:topic_kg}, topic modeling is selected adaptively based on available computational resources; consequently, the granularity and semantic resolution of topic assignments may vary across executions, while remaining structurally compatible with downstream knowledge graph construction.

Topic assignments and probabilities derived from the previous stage are embedded directly into the graph structure, enabling topic-centric navigation and analysis.

This dynamic construction strategy contrasts with static scholarly knowledge graphs, which require global maintenance and often introduce substantial noise for focused analytical tasks. By restricting knowledge graph scope to the query context defined in Equation~\ref{eq:doc_set}, ISLE achieves scalability while preserving relational richness.

\subsubsection{Data Flow and Architectural Principles}

Across all stages, ISLE adheres to four guiding architectural principles:

\begin{enumerate}
    \item Query-conditioned execution, ensuring that all downstream artifacts reflect the user's immediate information need.
    \item Modularity, allowing retrieval, topic modeling, and knowledge graph construction components to evolve independently.
    \item Resource-aware adaptivity, enabling the system to balance computational cost and semantic fidelity.
    \item Scalable subgraph construction, avoiding the overhead of maintaining a monolithic global knowledge graph.
\end{enumerate}

These principles enable ISLE to deliver an integrated yet flexible end-to-end pipeline, supporting interactive scientific literature exploration across heterogeneous deployment environments.

\section{Evaluation}\label{sec4}

This section provides a concise evaluation of ISLE, focusing on corpus characteristics, retrieval behavior, and an end-to-end qualitative case study. Given the system-oriented nature of ISLE, the evaluation emphasizes scalability, interpretability, and practical utility, rather than exhaustive benchmark optimization.

\subsection{Dataset and Experimental Context}

ISLE operates on a large-scale scholarly corpus constructed from arXiv and OpenAlex, comprising approximately 1.73 million papers spanning publications from 1939 to 2025. The corpus includes over 1.6 million distinct authors, 36,531 institutions, and 218 countries, with more than 3.4 million directed citation links. The average number of citations per paper is 11.75, reflecting a realistic and heterogeneous scientific landscape suitable for evaluating large-scale retrieval, topic modeling, and graph-based analysis.

All experiments were conducted in a query-driven setting, where retrieval, topic modeling, and knowledge graph construction are performed dynamically based on user input.

\subsection{Retrieval and Topic Modeling Behavior}

ISLE employs a hybrid retrieval strategy combining lexical and semantic search with Reciprocal Rank Fusion, as described in Section~3.2. While no manually curated relevance judgments were available for large-scale evaluation, the system behavior aligns with established findings that hybrid retrieval improves robustness across diverse query types \cite{ajith2024litsearch, bruch2023fusion, elastic2023hybrid, opensearch2024hybrid}.

For topic modeling, ISLE adopts a resource-aware strategy (Section~3.3), selecting between NMF-based and neural BERTopic-based pipelines depending on available computational resources. In practice, the neural pipeline consistently produced higher topic coherence and more semantically granular topic groupings, while the NMF-based approach offered a computationally efficient alternative with competitive interpretability.

\begin{figure}[h]
    \centering
    \includegraphics[width=\textwidth]{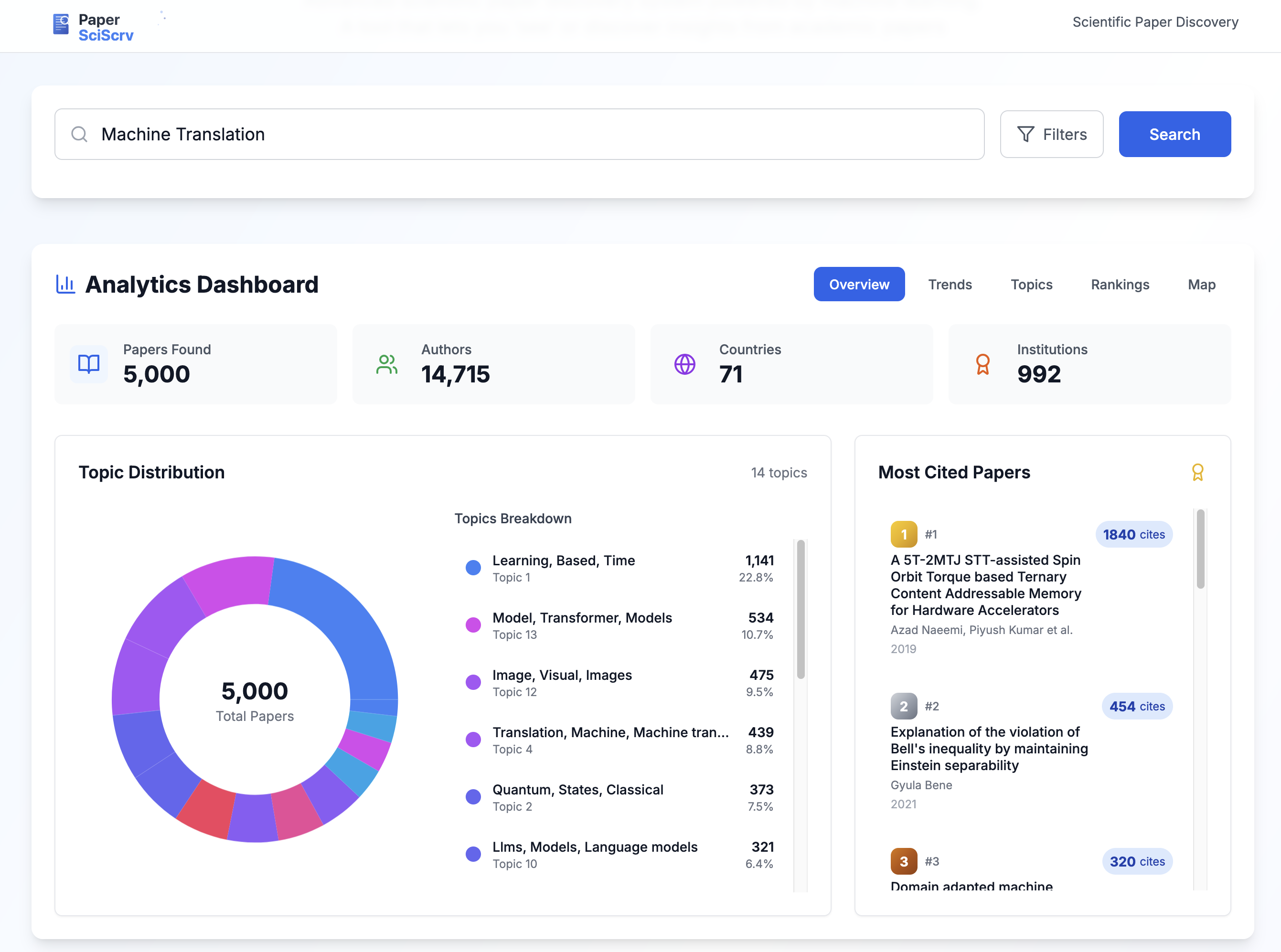}
    \caption{Analytics dashboard and topic distribution for the machine translation query}
    \label{fig:analytics}
\end{figure}

\subsection{Case Study: Machine Translation}

To illustrate ISLE's end-to-end capabilities, we conducted a qualitative case study using the query ``machine translation'', with a maximum retrieval limit of 5,000 papers. This query represents a well-established yet rapidly evolving research area, making it suitable for evaluating thematic structuring and relational analysis.

From the retrieved document set, ISLE constructed a query-conditioned knowledge graph containing:

\begin{itemize}
    \item 20,792 nodes and 224,521 edges,
    \item 14 latent topics,
    \item 14,715 authors,
    \item 992 institutions, and
    \item 71 countries.
\end{itemize}

\begin{figure}[htbp]
    \centering
    \includegraphics[width=\textwidth]{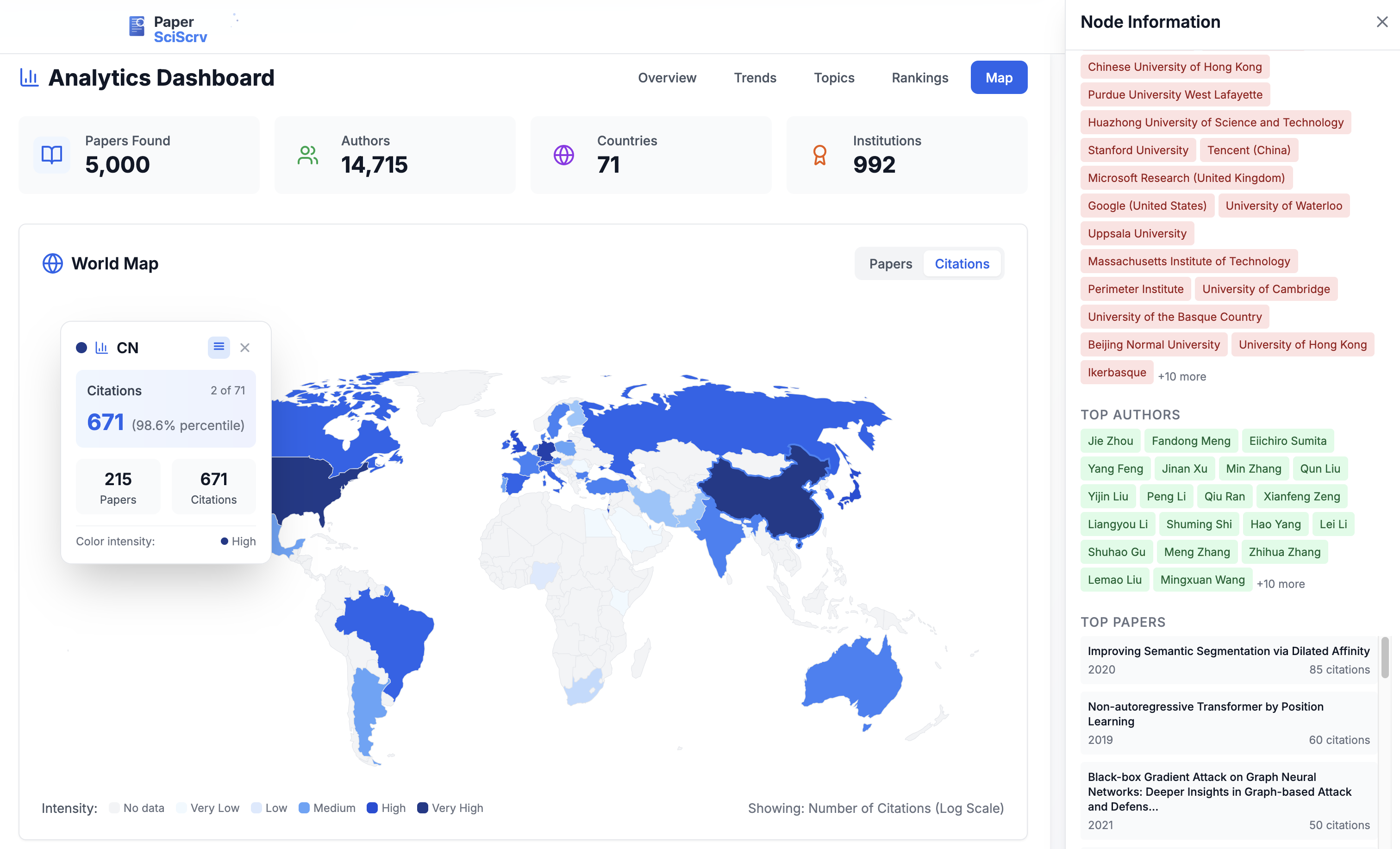}
    \caption{Citation-weighted geographic and institutional visualization}
    \label{fig:geographic}
\end{figure}

Figure~\ref{fig:analytics} presents the analytics dashboard for this query, summarizing the distribution of papers, authors, institutions, and countries. The topic distribution reveals clear thematic separation between foundational machine translation research, transformer-based models, multilingual learning, and related language modeling techniques.

Figure~\ref{fig:geographic} visualizes the geographic and institutional dimensions of the resulting graph, highlighting dominant research hubs and cross-country collaboration patterns. The citation-weighted world map enables rapid identification of regions with high scholarly impact in machine translation research.

\begin{figure}[h]
    \centering
    \includegraphics[width=\textwidth]{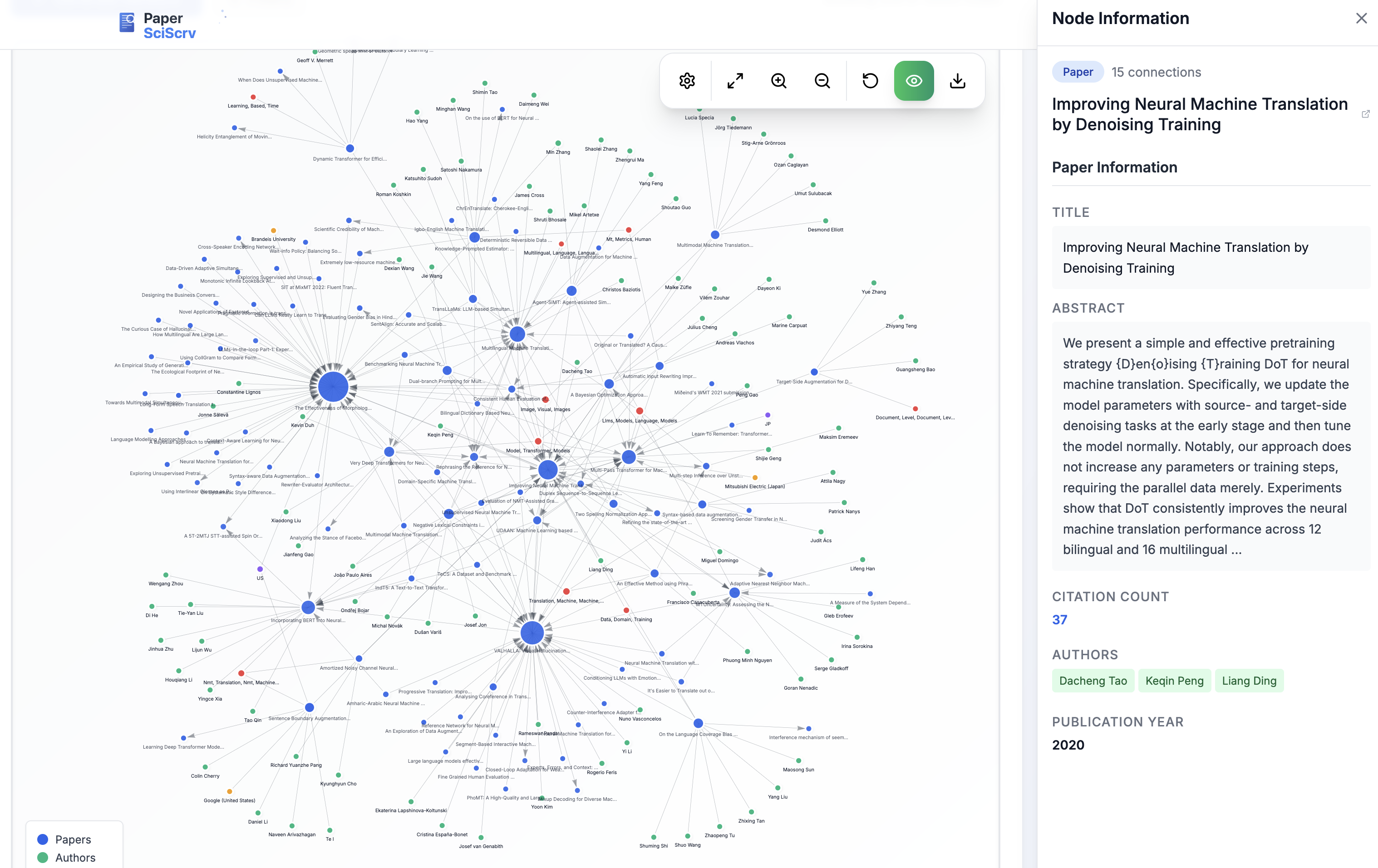}
    \caption{Query-conditioned knowledge graph visualization}
    \label{fig:kg_viz}
\end{figure}

Figure~\ref{fig:kg_viz} illustrates a subgraph-level visualization of the constructed knowledge graph, where papers, authors, institutions, locations, and topics are jointly represented. The knowledge graph structure exposes influential papers and authors as high-degree nodes, while topic nodes act as semantic anchors that facilitate exploration across related research clusters. Interactive inspection of individual nodes enables fine-grained analysis of publication metadata, citation counts, and authorship relationships.

\begin{figure}[h]
    \centering
    \begin{minipage}[b]{0.48\textwidth}
        \centering
        \includegraphics[width=\textwidth]{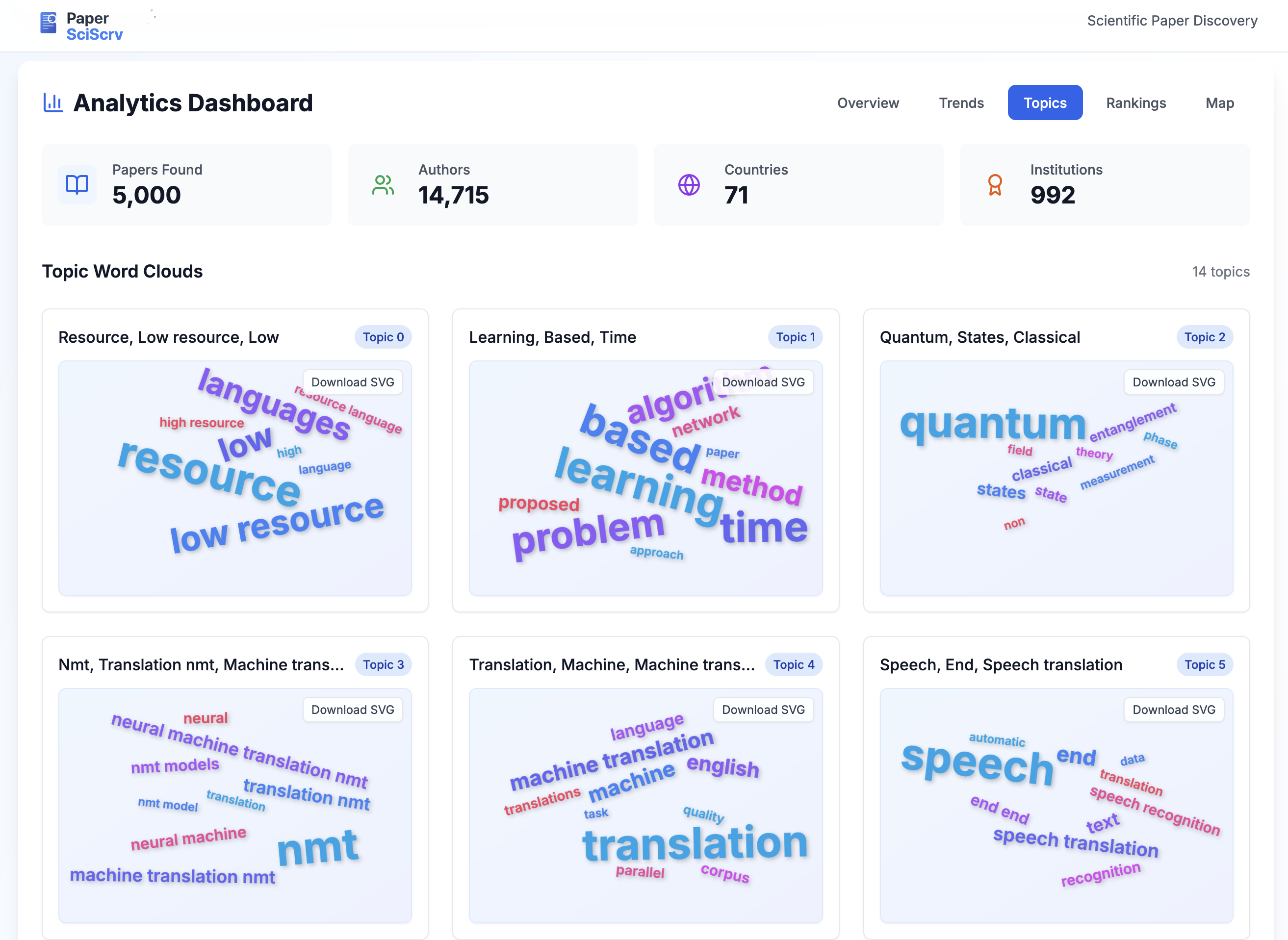}
    \end{minipage}
    \hfill
    \begin{minipage}[b]{0.48\textwidth}
        \centering
        \includegraphics[width=\textwidth]{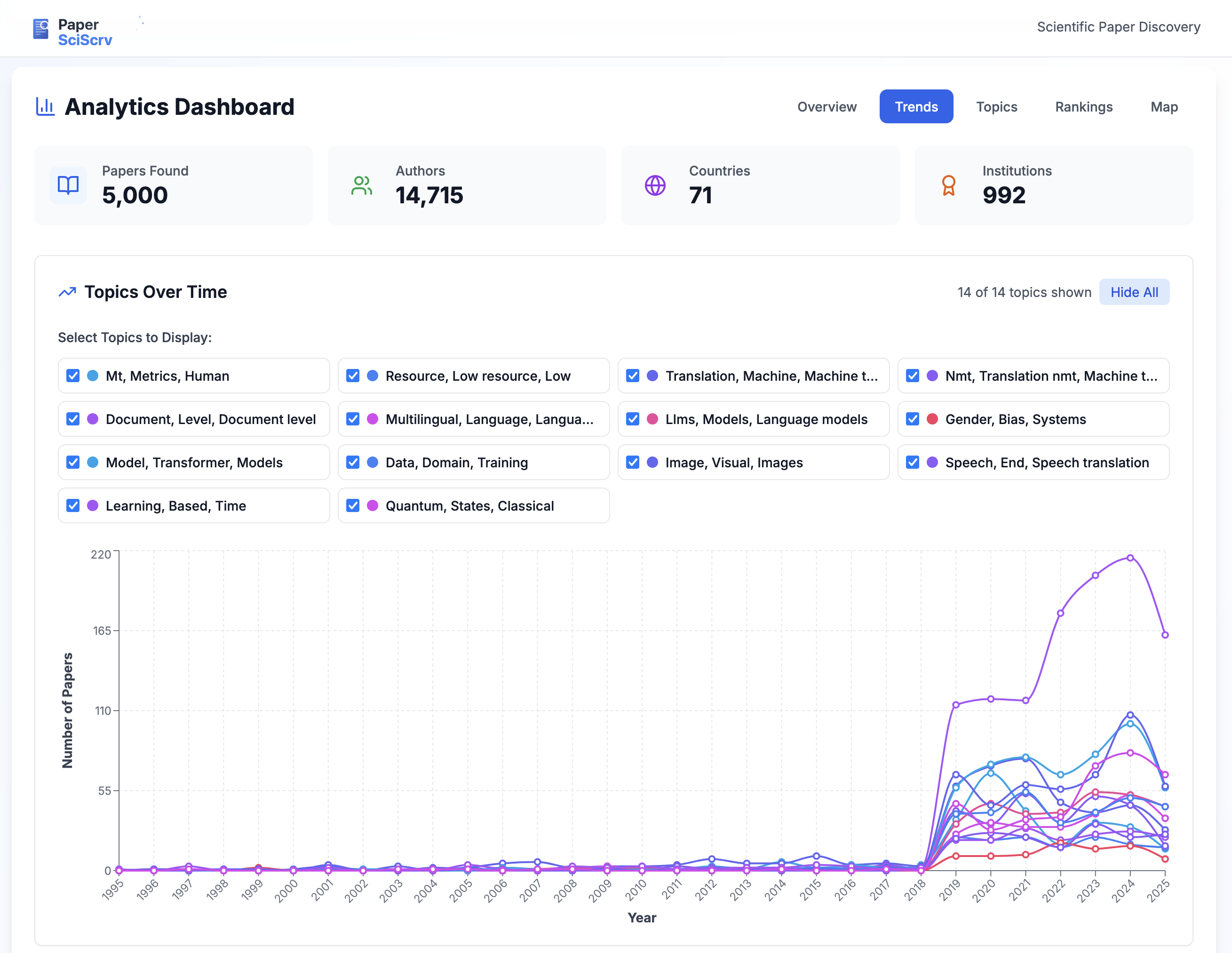}
    \end{minipage}
    \caption{Left: Query-conditioned knowledge graph visualization. Right: Topic word clouds for the machine translation query}
    \label{fig:combined}
\end{figure}

Figure~\ref{fig:combined} Left illustrates the semantic structure of the retrieved document set through topic-level word clouds generated by the topic modeling stage. Each topic is characterized by a compact set of high-weight terms, revealing clear thematic separation between core machine translation research (e.g., neural machine translation, transformers, low-resource languages) and related areas such as speech translation, multilingual learning, and evaluation. This visualization supports rapid qualitative assessment of topic coherence and interpretability.

Finally, Figure~\ref{fig:combined} Right presents the temporal distribution of publications across all identified topics. The plot reveals a pronounced growth in most topics after 2018, particularly for transformer-based models, large language models, and speech translation, reflecting recent shifts in machine translation research. In contrast, foundational and evaluation-oriented topics exhibit more gradual and stable trends over time.

\subsection{Summary}

This evaluation demonstrates that ISLE can efficiently construct large, semantically structured, and interpretable query-conditioned knowledge graphs from thousands of retrieved papers. By combining hybrid retrieval, resource-aware topic modeling, and dynamic knowledge graph construction, ISLE supports exploratory scientific analysis at scale while avoiding the complexity and noise of static global knowledge graphs.

\section{Discussion and Conclusion}\label{sec5}

This work presented ISLE, a query-conditioned scientific literature exploration system that integrates hybrid retrieval, resource-aware topic modeling, and dynamic knowledge graph construction into a unified analytical pipeline. Unlike conventional scholarly search engines or static knowledge graphs, ISLE is designed around the premise that scientific exploration is inherently contextual and query-conditioned, requiring representations that adapt to the user's information need rather than imposing a fixed global structure.

\subsection{Discussion}
A central design choice in ISLE is the use of hybrid retrieval, combining lexical and semantic signals through Reciprocal Rank Fusion. This approach enables the system to simultaneously capture precise terminological matches and deeper conceptual relevance, which is particularly important in scientific domains characterized by specialized vocabulary and rapid methodological evolution. The retrieval component serves not only as a ranking mechanism but also as a semantic filter that defines the scope of downstream analysis, directly shaping the structure of the constructed knowledge graph.

The resource-aware topic modeling strategy further strengthens ISLE's practical applicability. By dynamically selecting between computationally efficient matrix-factorization--based models and neural embedding--based approaches, the system maintains flexibility across deployment environments while preserving semantic consistency in downstream representations. Topic assignments are embedded directly into the knowledge graph structure, enabling topic-centric navigation, temporal analysis, and relational exploration within a single representational framework.

The query-conditioned knowledge graph constitutes the core analytical artifact produced by ISLE. In contrast to static scholarly knowledge graphs that require continuous global maintenance and often introduce noise for focused analytical tasks, ISLE constructs a typed, directed, and attributed multigraph tailored to each query context. This enables scalability to large corpora while preserving relational richness and interpretability. The resulting knowledge graphs support multi-level analysis---spanning papers, authors, institutions, countries, topics, and citations---without overwhelming the user with irrelevant entities.

\subsection{Limitations}

Despite its advantages, ISLE has several limitations. First, large-scale quantitative evaluation of retrieval effectiveness is constrained by the absence of comprehensive, query-level relevance judgments. Second, topic modeling quality remains sensitive to the size and thematic diversity of the retrieved document set, particularly for narrowly scoped queries. Finally, while the dynamic knowledge graph construction strategy improves scalability, interactive visualization of very large knowledge graphs may require additional abstraction or summarization mechanisms.

\subsection{Conclusion}

In summary, ISLE demonstrates that query-conditioned integration of retrieval, topic modeling, and knowledge graph construction provides a powerful paradigm for scientific literature exploration. By aligning analytical representations with user intent and computational constraints, ISLE enables scalable, interpretable, and semantically rich analysis of large scholarly corpora. Future work will explore personalization, citation-aware embeddings, and advanced knowledge graph summarization techniques to further enhance interactive scientific discovery.

\end{document}